\documentclass[aps,prd,twocolumn,showpacs,superscriptaddress]{revtex4}
\usepackage{bm}
\newcommand{\beq}{\begin{equation}}
\newcommand{\eeq}{\end{equation}}

\begin{document}

\title{Spatial entanglement and massive neutrino oscillations produced by
orbital electron capture decay}
\author{I.~M.~Pavlichenkov}
\email{pavi@mbslab.kiae.ru} \affiliation{Russian Research Center
Kurchatov Institute, Moscow, 123182, Russia}

\date{\today}

\begin{abstract}
The two-particle wave function of neutrino and recoil nucleus is
found as a solution of an initial value problem in the far zone
for a time longer than the electron capture decay lifetime of a
hydrogenlike ion. The neutrino-recoil entanglement arising in such
a process is a consequence of the momentum conservation and is
closely related to the wave packet structure of the state. Because
of neutrino mixing, the joint wave packet involves the coherent
superposition of the neutrino mass eigenstate packets. This is the
new physical realization of the Einstein-Podolsky-Rosen thought
experiment, which has no analogue in quantum optics and quantum
informatics. A class of possible experiments for the registration
of a neutrino and a recoil nucleus is proposed. It is shown that
due to spatial correlations neutrino oscillations can be observed
in the coincidence experiment with the recoil.

\end{abstract}

\maketitle

\noindent
\section{Introduction}

Massive neutrino oscillations are a consequence of the presence of
flavor neutrino mixing and a clear evidence of physics beyond the
standard model. They are subject of an intense experimental and
theoretical research beginning with the pioneering paper by
Pontecorvo \cite{Pont}. The theoretical approach is mostly
phenomenological and uses the plane wave \cite{Bil,Naka,Kay1} or
the wave packet formalism \cite{Nus,Kay,Giu,Dol,Giu1} to describe
the evolution of the massive neutrino states, and the
field-theoretical approach \cite{Grim,Grim1,Wu} which takes into
account the processes of production, propagation and detection of
neutrino. In spite of the fact that almost all these treatments
provide the canonical formula for the probability of oscillations,
some basic issues of the theory of neutrino oscillations are still
being debated \cite{Akh}. In particular, there is no consensus on
whether the three neutrino mass eigenstates have equal energies or
equal momenta. Furthermore, the wave packet formalism leaves
unanswered the question of how the properties of a neutrino wave
packet are determined by the process of weak decay.

If we want to resolve these problems, we need to do away with {\it
ad-hoc} assumptions and invoke basic principles of quantum
mechanics to describe the weak decay of an unstable object and
neutrino production \cite{Nau}. Closely related problems have been
studied in quantum optics. Fedorov {\it et al.} \cite{Fed,Fed1}
considered the decay of composite objects into two fragments that
are free to move away from the breakup point and are constrained
only by momentum and energy conservation. The authors obtained
position dependent two particle wave function as the solution of
an initial-value problem. This quantum state is entangled and
closely related to the Einstein-Podolsky-Rosen one \cite{Ein}.

Our motivation for this work is to apply the formalism of
Refs.~\cite{Fed,Fed1} to the description of orbital electron
capture (EC) decay. In this scenario, the spontaneous emission of
a flavor neutrino by an atom is described in three dimensions with
initial wave function of a decaying atom taken in the form of a
finite-size wave packet. The recoil-neutrino wave function is
found analytically in the coordinate representation as the
solution of the time-dependent Schr\"{o}dinger equation in a far
zone. Its wave packet structure evolves with time. The function
does not factorize in the neutrino, $\mathbf{r}_{\nu}$, and
recoil, $\mathbf{r}_n$, spatial coordinates, which is a direct
indication that the quantum state of the system is entangled.
However, due to neutrino mixing the entanglement involves the
coherent superposition of mass eigenstates with {\it equal
energies and different momenta} that manifests itself in the
oscillating correlations of the two particles.

Entanglement means that knowledge of one of the particles reveals
information about the other. Space correlations of neutrino and
recoil nucleus are the experimental implication of this feature.
It should be emphasized that the {\it spatial entanglement} which
we are considering, has nothing to do with {\it kinematic
entanglement} (see, for example, Refs.~\cite{Kay1,Gold}) which
does not involve an exact solution of the Schr\"{o}dinger
equation. Spatial entanglement between an atom and a photon in
spontaneous emission has been observed by Kurtsiefer {\it et al.}
\cite{Kurt}.

Because of strong spatial correlations, we are able to observe
with two detectors neutrino oscillations and a recoil
simultaneously in the coincidence measurement. This experiment,
first considered by Dolgov {\it et al.} \cite{Dol1}, could be of
interest for the next generation of neutrino detectors. By
contrast, registration of one particle independent of the other
disentangles the recoil-neutrino pair. For example, to observe
neutrino oscillations the recoil position is ignored whereas a
flavor neutrino is detected. This scenario is used currently to
observe neutrino oscillations in reactor and accelerator
experiments.

The paper is organized as follows. In Sec.~II we apply the
Weisskopf-Wigner theory of spontaneous emission \cite{Wei} to
treat the EC-decay process of the ion confined to a small volume
in a trap. We specifically study a hydrogen-like (H-like) ion of
the intermediate mass region with a K-electron in the lowest
hyperfine state having the total angular momentum $F = I\pm 1/2$,
where $I$ is the nuclear spin. Such ions have been used in the
much-talked-of GSI experiment \cite{Lit} and its second run in
2010 \cite{Lit1}. However, the question of what happens to the
neutrino and the recoil nucleus on time scales comparable to the
lifetime of the parent ion is left unanswered. The main result of
this section is the recoil-neutrino wave function that describes
the spatiotemporal behavior of these particles. In Sec.~III we use
the far zone approximations to express this function as a product
of the relative and the center of mass (CM) wave functions. It is
shown that the coherent superposition of the neutrino mass
eigenstates has a fixed kinetic energy equal to the EC-decay
energy. The structure and time evolution of the relative motion
(RM) and the CM wave packets are investigated in Sec.~IV. The far
zone formalism allows to find the spreading of these wave packets.
In Sec.~V we outline the experiments appropriate for the detection
of a flavor neutrino and a recoil nucleus. Our findings are
summarized in Sec.~VI. The Appendix contains the details of
calculation of an integral from Sec.~III.

\noindent
\section{Orbital electron capture decay of moving hydrogen-like ion}

The decay which we study is the capture of a single electron from
the K-shell into a completely ionized daughter nucleus and a
electron neutrino, as final state. The system is described by the
Hamiltonian $H=H_0+V_{sf}+H_w$, where the unperturbed Hamiltonian
has the form
\begin{equation}\label{hm0}
H_0=\!\frac{\hat\mathbf{p}^2}{2M}+H_n({\bm\rho}_1,...,{\bm\rho}_A)+
\!\sum_\lambda\varepsilon_\lambda a^+_\lambda a_\lambda+
\!\sum_{i,\mathbf{k}}\epsilon_i(k)c^+_{i\mathbf{k}}c_{i\mathbf{k}}.
\end{equation}
Here $\hat\mathbf{p}$ is the nucleus momentum operator; $H_n$ is
the nuclear hamiltonian depending on nucleon coordinates,
${\bm\rho}_k=\mathbf{r}_k-\mathbf{r}_n$, with respect to
$\mathbf{r}_n$ which to a good approximation is the nucleus CM;
$A$ is the mass number; $a^+_\lambda$ and $a_\lambda$ are the
creation and the annihilation operators of a bound electron in the
state with the energy $\varepsilon_\lambda$ and the quantum
numbers $\lambda=n,j,m$; $V_{sf}$ is a hyperfine interaction; and
$c^+_{i\mathbf{k}}$ and $c_{i\mathbf{k}}$ are the creation and the
annihilation operators of the three massive neutrinos ($i=1,2,3$)
with momentum $\mathbf{k}$, the energy
$\epsilon_i=\sqrt{k^2+m^2_i}$ and the mass $m_i$. The mass $M$ of
parent and daughter particles is supposed to be equal with an
accuracy of the small parameter $Q_{EC}/M$, where $Q_{EC}$ is the
decay energy.

The weak-interaction Hamiltonian density is given by
\begin{equation}\label{hmW}
h_w(\mathbf{r})=\frac{G_F}{\sqrt{2}}V_{ud}\left[\,\mathbf{j}(\mathbf{r})\!
\cdot\!\mathbf{J}(\mathbf{r})+{\rm H.c.}\right],
\end{equation}
where the lepton current,
\begin{equation}\label{wcur}
\mathbf{j}(\mathbf{r})=i\bar{\nu}_e(\mathbf{r}){\bm\gamma}
(1+\gamma_5)e(\mathbf{r}),
\end{equation}
involves the flavor neutrino and electron field operators. The
former is \cite{vol}
\begin{equation}
\nu_e(\mathbf{r})=\sum_iU_{ei}\nu_e(\mathbf{r})=
\sum_{i,{\mathbf{k}_i}}U_{ei}
c_{i{\mathbf{k}_i}}u_L(\mathbf{n}_i)e^{i\mathbf{k}_i\mathbf{r}},
\end{equation}
where $U$ is the Pontecorvo-Maki-Nakagava-Sakata mixing matrix.
The Dirac spinor for left-handed neutrino,
\begin{equation}
u_L(\mathbf{n})\!=\!\frac{1}{\sqrt{2}}\left(\!\omega_L\atop
-\omega_L\!\right),\
\omega_L(\mathbf{n})\!=\!\left(-e^{-i\varphi/2}\sin{(\theta/2)}\atop
e^{i\varphi/2}\cos{(\theta/2)}\right),
\end{equation}
depends on the unit vector $\mathbf{n}=\mathbf{k}/k$. We will use
non-relativistic limit for the electron operator
\begin{equation}
e(\mathbf{r})=\sum_\lambda a_\lambda\psi_\lambda(\mathbf{r}),\ \
\psi_\lambda=\left(\!w_\lambda\atop 0\!\right). \vspace{-2mm}
\end{equation}
For the K-electron, we have
$w_{1\frac{1}{2}m}=f_{1\!s}(r)\eta_{\frac{1}{2},m}$, where $f$ is
radial and $\eta$ is spin wave functions. The nucleon current for
pure Gamow-Teller transitions in a non-relativistic approximation
has the form
\begin{equation}\label{ncur}
\mathbf{J}(\mathbf{r})\!=\!g_A\!\sum^A_{s=1}{\bm
\sigma}(s)\tau_-(s)\delta(\mathbf{r}\!-\!\mathbf{r}_s)\!=\!
g_A\!\sum^A_{s=1}\mathbf{J}(s)\delta(\mathbf{r}\!-\!\mathbf{r}_s),
\end{equation}
where $\sigma$ and $\tau$ are the spin and the isospin Pauli matrices.

To find the wave packet structure of an entangled state, one needs
to use the coordinate representation for the wave functions of
neutrino and recoil nucleus. This implies that the Hamiltonian
$H_w$ has to commute with the total momentum of a bipartite state
in the $\mathbf{r}$-representation. By using Eqs.~(\ref{hmW}),
(\ref{wcur}) and (\ref{ncur}) we get~after integration over
nuclear volume the weak interaction Hamiltonian fulfilling the
momentum conservation law
$$
H_w=i\frac{G_F}{\sqrt{2}}V_{ud}\sum_{i,{\mathbf{k}_i}}
U^*_{ei}c^+_{i{\mathbf{k}_i}}\bar{u}_L(\mathbf{n}){\bm\gamma}(1+\gamma_5)
$$
\vspace{-5mm}
\begin{equation}\label{hmWS}
\times\!\sum_s\mathbf{J}(s)e^{-i\mathbf{k}_i{\bm\rho}_s}
\sum_\lambda a_\lambda\psi_\lambda({\bm\rho}_s)
e^{i\mathbf{k}_i(\mathbf{r}_\nu-\mathbf{r}_n)} +{\rm H.c.}.
\vspace{-2mm}
\end{equation}

Now let us use the time-dependent perturbation theory to determine
the temporal evolution of a decaying state for the Gamow-Teller
transition $FIM_F\rightarrow I'M'$ with $I'=I\pm 1$. The parent
ion is prepared in a polarized state with the angular momentum
projection $M_F$ along the quantized axis $Z$ of the laboratory
frame $S$. The solution of the Schr\"{o}dinger equation with the
Hamiltonian $H$ is sought by using the following ansatz for the
wave function
$$
\Psi(t)=\sum_\mathbf{p}{\cal A}(\mathbf{p},t)\left|
aFIM_F;\mathbf{p},t\right>e^{-i{\cal E}_at}
$$
\vspace{-6mm}
\begin{equation}\label{anz}
+\sum_{i,M',\mathbf{p},{\mathbf{k}_i}}\!\!\!\!{\cal
B}_{iM'}(\mathbf{p},\mathbf{k}_i,t)
\left|bI'M';\mathbf{p}-\mathbf{k}_i,\mathbf{k}_i,t\right>e^{-i{\cal
E}_{bi}t},
\end{equation}
where
$$
\left|aFIM_F;\mathbf{p}\right>=\!
\sum_{m,M}C^{FM_F}_{\frac{1}{2}m,I\!M}
\Omega_{aIM}({\bm\rho})a^+_{1\frac{1}{2}m}\left|0\right>
e^{i\mathbf{p}\mathbf{r}_n},
$$
\vspace{-6mm}
\begin{equation}\label{eigf}\hspace{-0.6pt}
\left|bI'\!M';\mathbf{p}-\mathbf{k}_i,\mathbf{k}_i\right>\!= \!
\Omega_{bI'\!M'}({\bm\rho})\,c^+_{i\mathbf{k}_i}\!\left|0\right>
e^{i(\mathbf{p}-\mathbf{k}_i)\mathbf{r}_n+i\mathbf{k}_i\mathbf{r}_\nu}
\end{equation}
are the eigenvectors of $H_0+V_{sf}$ with the eigenvalues
\begin{equation}
{\cal E}_a=E_{aI}\!+\!\varepsilon_0\!+\!\frac{\mathbf{p}^2}{2M},\
\ {\cal E}_{bi}=E_{bI'}\!+\!\epsilon_i\!+
\!\frac{(\mathbf{p}-\mathbf{k}_i)^2}{2M}.
\end{equation}
In Eq.~(\ref{eigf}) $\Omega({\bm\rho})=\Omega({\bm\rho}_1,...,{\bm\rho}_A)$ and
$E$ are eigenfunctions and eigenvalues of the initial ($aI$) and
final ($bI'$) states of nucleus, $\varepsilon_0$ is the energy of
the ion ground state, $\left|0\right>$ is the lepton vacuum, and
$C^{FM_F}_{1/2m,IM}$ are Clebsch-Gordan coefficients. In the
subsequent text we will assume that $E_{bI'}=0$.

The differential equations for the coefficients ${\cal A}$ and
${\cal B}$ are
\begin{eqnarray}\label{eqa}
&i\!\!\!&\dot{\cal A}(\mathbf{p},t)=\!\!\sum_{i,{\mathbf{k}_i},M'}
\!\!\!{\cal
B}_{iM'}(\mathbf{p},\mathbf{k}_i,t)W_{I'\!M'}(\mathbf{n}_i)U_{ei}
e^{i({\cal E}_a-{\cal E}_{bi})t},\nonumber\\
&i\!\!\!&\dot{\cal B}_{iM'}(\mathbf{p},\mathbf{k}_i,t)={\cal
A}(\mathbf{p},t)W^*_{I'\!M'}(\mathbf{n}_i)U^*_{ei} e^{-i({\cal
E}_a-{\cal E}_{bi})t},
\end{eqnarray}
where the matrix element of the Hamiltonian $H_w$,
\begin{equation}\label{W}
W_{I'\!M'}(\mathbf{n}_i)\!=\!\frac{G_FV_{ud}}{\sqrt{F\!+\!1/2}}
\!\left<f_{1s}\right>\! g_A{\cal
M}(aI,\!bI')\xi^*_{I'\!M'}(\mathbf{n}_i\!),
\end{equation}
includes the radial wave function of the bound electron averaged
over nuclear volume, $\left<f_{1s}\right>$, and the nuclear
reduced matrix element ${\cal M}$. The spinor
\begin{equation}
\xi_{I'\!M'}(\mathbf{n}\!)= C^{FM_F}_{I'M',\,\frac{1}{2}\mu}
D^{1/2}_{\mu,-\frac{1}{2}}(\varphi,\theta,0),
\end{equation}
where $D$ is the Wigner function and $\mu=M_F-M'=\pm\frac{1}{2}$,
is normalized by the condition
$\sum_{M_F,M'}\!|\xi_{I'\!M'}|^2\!=\!F\!+\!1/2$.

Suppose that a parent ion after production is stored in a trap and
its motion is localized. Then we switch off the field of the trap
and free spreading of the ion CM wave packet begins. If the
production and stored time is far less than the life-time of the
parent ion, we can measure time from the beginning of free
spreading. The initial state is described by the first term of
Eq.~(\ref{anz}) for $t=0$. To be specific, suppose that the CM
part of this function has in the momentum representation the
Gaussian form
\begin{equation}\label{init}
{\cal A}(\mathbf{p},t\!=\!0)\!=\!
(2\sqrt{\pi}d)^{3/2}\exp\left(\!-\frac{1}{2}d^2p^2\right).
\end{equation}
The Eqs.~(\ref{eqa}) are easily solved in the Weisskopf-Wigner
approximation with the initial conditions ${\cal
A}(\mathbf{p},t=0)={\cal A}_0(\mathbf{p})$ and ${\cal
B}_{iM'}(\mathbf{p},\mathbf{k},t=0)=0$. The solutions for $t\gg
1/Q_{EC}$ are
$$
{\cal A}(\mathbf{p},t)={\cal A}_0(\mathbf{p})\exp\left(-\Gamma
t\right),
$$
\vspace{-6mm}
\begin{equation}\label{coef}
{\cal B}_{iM'}(\mathbf{p},\mathbf{k},t)\!=\!\frac{{\cal
A}_0(\mathbf{p})W^*_{I'\!M'}(\mathbf{n}_i)U^*_{ei}}{{\cal
E}_{bi}-{\cal E}_a+i\Gamma}\!\left[1\!-\!e^{i({\cal E}_{bi}-{\cal
E}_a)t-\Gamma t}\right],
\end{equation}
 where\vspace{-3mm}
$$
\Gamma=\pi\!\!\!\!\sum_{i,M',\mathbf{k}_i}\!\!\!\!
\left|W_{I'\!M'}(\mathbf{n}_i)\right|^2\left|U_{ei}\right|^2\delta({\cal
E}_{bi}-{\cal E}_{a})
$$
\vspace{-5mm}
\begin{equation}\label{gam}
=\frac{(G_FV_{ud})^2}{2\pi(2F+1)}[\left<f_{1s}\right>\!g_A{\cal
M}(aI,\!bI')Q_{EC}]^2
\end{equation}
is one half of the rate of the Gamow-Teller transition
$FIM_F\rightarrow I'M', I'=I\pm 1$. The final expression we have
derived for this value ignores a small dependence of $\Gamma$ on
the parent ion velocity. Furthermore, the two small dimensionless
parameters
\begin{equation}\label{smal}
\alpha=\frac{Q_{EC}}{M},\quad \delta_i=\frac{m_i}{Q_{EC}}
\end{equation}
 allow to
reduce $\Gamma$ to a conventional value for the ion at rest. The
last parameter is compatible with the limit of ultra relativistic
neutrinos \cite{Giu2}.

First of all, we observe that the decay probability $P(t)$ of the
parent ion is not dependent on the shape of the initial CM wave
packet ${\cal A}_0(\mathbf{p})$ because of
\begin{equation}
P(t)=\exp{(-2\Gamma t)}\!\sum_\mathbf{p}\!{\cal
A}^2_0(\mathbf{p})=\exp{(-2\Gamma t)}.
\end{equation}
However the probability of emission of the electron neutrino with
wave vector $\mathbf{k}$ and energy $\epsilon$ is little affected
by Doppler shift
\begin{equation}
P_e(\mathbf{k})\!=\!\sum_{M'}\left|W_{I'\!M'}(\mathbf{n})\right|^2\!
\sum_\mathbf{p}\frac{{\cal
A}^2_0(\mathbf{p})}{(\epsilon\!-\!Q_{EC}\!+\!\frac{\mathbf{k}^2}{2M}\!
-\!\frac{\mathbf{p}\cdot\mathbf{k}}{M})^2\!+\!\Gamma^2}.
\end{equation}

With ${\cal B}_{iM'}$ taken from Eq.~(\ref{coef}), the
recoil-neutrino wave function at times $t>1/\Gamma$ has the
form
$$
\Psi_{I'\!M'}(\mathbf{r}_n,\mathbf{r}_\nu,t)=
\!\!\sum_{i,\mathbf{p},\mathbf{k}_i}\frac{{\cal A}_0(\mathbf{p})
W^*_{I'\!M'}(\mathbf{n}_i)}
{\epsilon_i\!-\!Q_{EC}\!+\!\frac{\mathbf{k}^2_i}{2M}\!
-\!\frac{\mathbf{p}\mathbf{k}_i}{M}+i\Gamma}
$$
\vspace{-4mm}
\begin{equation}\label{func}
\times\exp\!\left\{i(\mathbf{p}\!-\!\mathbf{k}_i)\mathbf{r}_n\!+
i\mathbf{k}_i\mathbf{r}_\nu\!-\!i{\cal
E}_{bi}t\right\}U^*_{ei}c^+_{i\mathbf{k}_i}\!\left|0\right>.
\end{equation}
It carries information on the neutrino production process and
involves the coherent superposition of the electron neutrino mass
eigenstate components.

\noindent
\section{Recoil-neutrino wave function}
Now, by using the far zone approximation and small parameters
(\ref{smal}), we write the function (\ref{func}) in the form
suitable for the analysis of the wave packet structure and
entanglement of the recoil-neutrino state. To begin with, let us
rewrite Eq.~(\ref{func}) as
\begin{equation}\label{psi}
\Psi_{I'\!M'}\!=\! \!\sum_{\mathbf{p}}\! {\cal
A}_0(\mathbf{p})\exp\! \left(\!i\mathbf{p}\mathbf{r}_n\!-\!
i\frac{\mathbf{p}^2t}{2M}\right)\!
\sum_i\chi_i(\mathbf{r},t)\,U^*_{ei}c^+_{i\mathbf{k}_i}\!\!\left|0\right>,
\end{equation}
where the function
\begin{equation}\label{psii}
\chi_i=\!\!\sum_{\mathbf{k}_i}\!
\frac{W^*_{I'\!M'}(\mathbf{n}_i)\exp{\!\left[i\mathbf{k}_i\mathbf{r}-
i\left(\epsilon_i\!+\!\frac{\mathbf{k}^2_i}{2M}\!
-\!\frac{\mathbf{p}\mathbf{k}_i}{M}\right)\!t\right]}}
{\epsilon_i\!-\!Q_{EC}\!+\!\frac{\mathbf{k}^2_i}{2M}\!
-\!\frac{\mathbf{p}\mathbf{k}_i}{M}+i\Gamma}
\end{equation}
depends on the relative coordinate
$\mathbf{r}=\mathbf{r}_\nu-\mathbf{r}_n$. In the laboratory frame
$S$, the vector $\mathbf{r}$ is specified by polar angles $\theta$
and $\varphi$. The direction of $\mathbf{k}_i$ is given by angles
$\vartheta_i$ and $\phi_i$.

First we transform the summation over $\mathbf{k}_i$ in
Eq.~(\ref{psii}) to integration. To perform integration over
$d\Omega_{\mathbf{k}_i}$, we rotate the laboratory system $S$ with
the origin $O$ in the decay point by the angles $\theta$,
$\varphi$ to align axis $Z$ along the vector $\mathbf{r}$. In the
new frame $S'$, the vectors $\mathbf{k}_i$ and $\mathbf{p}$ are
determined by angles $\vartheta'_i,\phi'_i$ and
$\vartheta_p,\phi_p$, respectively. The polar angles of these
vectors in the systems $S$ and $S'$ are connected with each other
by well-known formulas~\cite{Var}. Owing to $k_ir\sim Q_{EC}t>
Q_{EC}/\Gamma\gg 1$, the integrant involves rapidly oscillating
function $\exp{(ik_ir\cos\vartheta'_i)}$ and a slowly varying
preexponential function ${\cal F}$ of the angle $\vartheta'_i$.
Integration by parts over this angle yields
$$
\int^\pi_0\!\!{\cal F}(\vartheta'_i)\exp{(ik_ir\cos\vartheta'_i)}
\sin\vartheta'_id\vartheta'_i
$$
\vspace{-5mm}
\begin{equation}
=\frac{i}{k_ir}\left[{\cal F}(\pi)e^{-ik_ir}-{\cal
F}(0)e^{ik_ir}\right]+{\cal O}(1/(k_ir)^2).
\end{equation}
Here the two terms in brackets represent incoming and outgoing
spherical waves. The incoming wave in the far zone gives an
exponentially small contribution for $t>1/\Gamma$ which can be
neglected in comparison with the outgoing one. The latter
corresponds to $\vartheta'_i=0$, and it immediately follows that
$\vartheta_i=\theta$ and $\phi_i=\varphi$ in the integrant of
(\ref{psii}). Thus, we find that $\mathbf{k}_i\|\mathbf{r}$ and
$\mathbf{n}_i=\mathbf{n}=\mathbf{r}/r$ in the far zone. After
integration over $d\phi'_i$, one obtains
\begin{equation}\label{psii1}
\chi_i=\! \frac{W^*_{I'\!M'}(\mathbf{n})}{i(2\pi)^2r}
\!\int\limits^\infty_0 \frac{\exp{\!\left[ik_ir\!-\!
i\left(\epsilon_i\!+\!\frac{k^2_i}{2M}\!
-\!\frac{\mathbf{p}\mathbf{k}_i}{M}\right)\!t\right]}}
{\epsilon_i\!-\!Q_{EC}\!+\!\frac{k^2_i}{2M}\!
-\!\frac{\mathbf{p}\mathbf{k}_i}{M}+i\Gamma}k_idk_i,
\end{equation}
where $\mathbf{p}\mathbf{k}_i=pk_i\cos{\vartheta_p}$.

To calculate the integral, we change variable $k_i$ by
$\epsilon_i$ and expand the former around $Q_{EC}$
\begin{equation}\label{exp}
k_i(\epsilon_i)=k_{i0}+\frac{1}{v_i}(\epsilon_i-Q_{EC})-
\frac{m^2_i}{2v^3_iQ^3_{EC}}(\epsilon_i-Q_{EC})^2,
\end{equation}
because main contribution to the pole in (\ref{psii1}) comes from
$\epsilon_i=Q_{EC}-i\Gamma$. We have retained a quadratic term to
describe the spreading of the RM wave packet. Here
$k_{i0}=\sqrt{Q^2_{EC}-m^2_i}$ and $v_i=k_{i0}/Q_{EC}$ is the
group velocity of neutrino. Then the integral over $\epsilon_i$
can be evaluated by the residue method. The details of
calculations are given in the Appendix. Upon using Eqs.~(\ref{W})
and (\ref{gam}), we find  in the lowest order in the parameters
$\alpha$ and $\delta_i$
\begin{equation}\label{chi}
\chi_i=-\psi_i(\mathbf{n},r,t)\exp(i\alpha\mathbf{p}\mathbf{r}),
\end{equation}
where the RM wave function of $i$th massive
neutrino is
$$
\psi_i(\mathbf{n},r,t)\!=\!\sqrt{\frac{\Gamma}{4\pi}}
\frac{1}{r}\left\{1\!-\!{\rm Erf}\!\left[\sqrt{\frac{i}{2}}
\left(\!\Delta_i+\!i\frac{v_it-r}{\Delta_iD_i}
\right)\right]\right\}
$$
\vspace{-3mm}
\begin{equation}\label{relat}
\times\xi_{I'\!M'}(\mathbf{n}\!)\exp{\left[i(k_{i0}r-Q_{EC}t)-
(v_it-r)/D_i\right]}.
\end{equation}
Here Erf is the error function, $D_i=v_i/\Gamma$ is the initial
width of the RM wave packet, and
\begin{equation}\label{delt}
\Delta_i=\frac{1}{D_i}\left(\frac{t}{M}+
\frac{\delta^3_ir}{m_iv_i}\right)^{1/2}
\end{equation}
is a dimensionless parameter describing its spreading.

Now, integration over $d\mathbf{p}$ can be readily performed after
substituting expressions (\ref{init}) and (\ref{chi}) into
Eq.~(\ref{psi})
$$
\sum_{\mathbf{p}}\!{\cal A}_0(\mathbf{p})\exp\!
\left[i\mathbf{p}(\mathbf{r}_n+\alpha\mathbf{r})\!-\!
i\frac{\mathbf{p}^2t}{2M}\right]=
$$
\vspace{-4mm}
\begin{equation}
=\frac{1}{\pi^{3/4} \left(d+\frac{it}{Md}\right)^{3/2}}
\exp{\!\left[-\frac{(\mathbf{r}_n+\alpha
\mathbf{r})^2}{2d\left(d+\frac{it}{Md}\right)}\right]}.
\end{equation}
In view of the definition of the CM vector for a relativistic
system in terms of the particle energies \cite{lanl},
\begin{equation}
\mathbf{R}=\mathbf{r}_n+\frac{\epsilon_i}{\epsilon_n+\epsilon_i}
\mathbf{r} \approx \mathbf{r}_n+\alpha\mathbf{r},
\end{equation}
we find an analytic expression for the wave function (\ref{psi})
$$
\Psi_{I'\!M'}(\mathbf{R},\mathbf{r},t)\!=\!-\frac{1}{\pi^{3/4}
\left(d+\frac{it}{Md}\right)^{3/2}}
\exp{\!\left[-\frac{\mathbf{R}^2}{2d\left(d+\frac{it}{Md}\right)}\right]}
$$
\vspace{-2mm}
\begin{equation}\label{final}
\times\sum_i\psi_i(\mathbf{n},r,t)U^*_{ei}c^+_{i\mathbf{k}0}
\!\left|0\right>.
\end{equation}\vspace{-4mm}

The two-particle wave function (\ref{final}) describes the
evolution of the recoil-neutrino state after decay at time
$t>1/\Gamma$. This expression is accurate to within the small
parameters $\alpha$ and $\delta_i$ (\ref{smal}). The function
carries information on the decay process and has the form of a
product of the CM and RM parts. Such a factorization of the total
wave function is a general feature of the two-particle decay of
noninteracting fragments constrained only by momentum and energy
conservation \cite{Fed1}. The RM wave function involves the
coherent superposition of the mass eigenstate components of the
electron neutrino state.

The time-dependent phase factor of the recoil-neutrino wave
function (\ref{final}) [see also Eq.~(\ref{relat})] involves the
kinetic energy of the pair. As shown in the Appendix, it is equal
with great accuracy to the decay energy $Q_{EC}$. Thus, all three
massive neutrinos have the same energy. This resolves the long
standing problem "same energy or same momentum." This issue, like
other paradoxes of neutrino oscillations \cite{Akh}, emerges from
the theory that considers the time evolution of a neutrino
independently from a recoil. The exact solution (\ref{final})
shows that a neutrino and a recoil do not evolve separately due to
their spatial correlation.

\noindent
\section{The wave packet structure of the recoil-neutrino state
and spatial entanglement}

The spatiotemporal behavior of the joint quantum state of the
recoil and the electron neutrino following the EC-decay is in
agreement with the results obtained in the theoretical studies of
decaying bipartite systems \cite{Fed,Fed1,Law}. The distinctive
feature of our system is the coherent superposition of neutrino
mass eigenstates. In this section we give the more detailed
analysis of the function (\ref{final}) providing insight into the
nature of entanglement and neutrino oscillations. We begin with
the wave packet structure of this function, since entanglement and
neutrino oscillations depend on spatial localization of particles
involved in the decay.

The CM part of $\Psi_{I'\!M'}(\mathbf{R},\mathbf{r},t)$ has the
form of a spreading wave packet
\begin{equation}\label{cm}
\mid\Psi_{CM}(\mathbf{R},t)\mid^2=\frac{1}{\pi^{3/2}D^3_R(t)}
\exp\left[-\frac{\mathbf{R}^2}{D^2_R(t)}\right],
\end{equation}
with the time-dependent width
\begin{equation}
D_R(t)=\!\left[d^2\!+\!\left(\frac{t}{Md}\right)^2\right]^{1/2}\!\!\!=\!
\left\{\begin{array}{cc} d,& t\ll t_{CM}\\
l^2_M/d,& t\gg t_{CM} \\\end{array}\right.,
\end{equation}
where $t_{CM}=Md^2$ is its spreading time and $l_M=\sqrt{t/M}$ is
the quantum diffusion length. If the initial size $d$ of the
atomic wave packet is approximately $10^{-6}$ cm, we have
$t_{CM}\sim 10^{-6}$ s for nucleus with the mass number $A\sim
100$. Hence, the width $D_R$ is of pure dispersion origin and
grows linearly with time. The velocity of spreading is equal to
$1/Md$.

The time-dependent width of the RM wave packet
$|\!\!~\psi_i(\mathbf{r},t)\!\!\mid^2$ is due to dispersive
broadening of both the recoil and the massive neutrino. To
estimate these effects, let us put $r\approx v_it$ in
Eq.~(\ref{delt}). We get
\begin{equation}
\Delta_i=\frac{1}{D_i}(l^2_M+\delta^3_il^2_{m_i})^{1/2}\approx
\frac{l_M}{D_i}=\sqrt{\frac{t}{t_i}},
\end{equation}
where $t_i=MD^2_i$ is the spreading time caused only by a recoil
particle. The contribution of a massive neutrino is negligible
because of the relativistic suppression of the wave packet
spreading. For $\Gamma\sim 1\,{\rm s}^{-1}$ and $A\sim 100$, the
spreading time is $t_i\sim 10^{19}$ yr. This result is apparent
for ultrarelativistic neutrinos, and we can put $\Delta_i=0$ in
Eq.~(\ref{relat}). In this limit, one finds the RM function
\begin{equation}\label{rm}
\psi_{I'\!M'}(\mathbf{n},r,t)\!=\!\frac{1}{\sqrt{2\pi}}
\xi_{I'\!M'}(\mathbf{n})\!\sum_i\!R_i(r,t)
U^*_{ei}c^+_{i\mathbf{k}_{i0}}\!\left|0\right>,
\end{equation}
where the radial function of the $i$th massive neutrino is
\begin{equation}\label{radf}
R_i(r,t)\!=\!\frac{\sqrt{2\Gamma}}{r}\exp{\!\!\left[i(k_{i0}r\!-\!
Q_{EC}t)\!-\!\frac{v_it\!-\!r}{D_i}\right]}\Theta(v_it\!-\!r).
\end{equation}
Here $\Theta$ is the unit step function. The function $R_i(r,t)$
is Lorentz covariant. The RM function (\ref{rm}) is normalized for
ultra relativistic neutrinos by the condition
\begin{equation}
\int d\mathbf{r}\sum_{M'}\mid\psi_{I'M'}(\mathbf{n},r,t)\mid^2=
\sum_i\mid U_{ei}\mid^2=1.
\end{equation}

The wave packet $\mid\!\psi_{I'M'}(\mathbf{r},t)\!\mid^2$ is a
superposition of three exponential wave packets of massive
neutrinos with different sharp edges $r=v_it$ and widths $D_i$.
The later depends only on the dynamics of the decay process. The
difference between the group velocities $v_i$ of these packets
results in their separation. However, the separation is negligible
compared with $D_i$ for times $t\ll 10^{12}$ yr. Therefore the
wave packets corresponding to different mass eigenstates are in
fact spatially inseparable. Hence, we can use an ultra
relativistic approximation for $R_i$ with $D_i=D=1/\Gamma$ and
$v_i=1$. In this approximation the function (\ref{radf}) takes the
form
\begin{equation}
R_i(r,t)\!=\!R_0(r,t)\exp[i(k_{i0}r\!-\! Q_{EC}t)],
\end{equation}
where
\begin{equation}
R_0(r,t)=\frac{1}{r}\sqrt{\frac{2}{D}}\exp\left(\!-\frac{t\!-\!r}{D}\right)
\Theta(t\!-\!r).
\end{equation}

We now wish to consider the two-particle wave function, taken as
the product of two functions (\ref{cm}) and (\ref{rm}), in the
observable coordinates of the recoil nucleus and the neutrino. This
normalized function has the form
\begin{equation}
\Psi_{I'\!M'}\!=\!
\Psi_{CM}\bigl((1-\alpha)\mathbf{r}_n\!+\!\alpha\mathbf{r}_\nu,t\bigl)
\psi_{I'\!M'}(\mathbf{n},|\mathbf{r}_\nu\!-\!\mathbf{r}_n|,t).
\end{equation}
The function does not factorize in these variables -- a direct
indication of the spatial entanglement of two particles. Each of
the three massive neutrinos becomes entangled with the recoil
nucleus because a neutrino is not emitted in a momentum
eigenstate.

The joint recoil-neutrino wave packet
$\sum_{M'}\left|\Psi_{I'\!M'}\right|^2$ has an axially symmetrical
shape with respect to the axis passing through the CM in the
direction of the vector $\mathbf{n}$. The packet increases with
time in a transverse direction with velocity $1/Md$ and in a
longitudinal (along the axis) one with velocity $v_n+(1-v_n)=1$,
where $v_n=v_i\epsilon_i/\epsilon_n\approx\alpha v_i$ is the
velocity of a recoil. It is easy to see that the function
$\Phi_{CM}$ takes its maximum value on the symmetry axis, along
which the state is highly entangled. The probability density is
proportional to a product of the Gaussian and exponential
functions
$$
\sum_{M'}\left|\Psi_{I'\!M'}\right|^2\sim
F_{I'}(\theta)\exp{\left\{\!-\frac{[(1-\alpha)r_n-\alpha
r_\nu]^2}{D^2_R}\right\}}\vspace{-4.5mm}
$$
\begin{equation}\label{paxis}
\times\exp{\left(\!-\frac{t-r_\nu-r_n}{D}\right)}\Theta(t-r_\nu-r_n),
\vspace{-1mm}
\end{equation}
where the angular modulation of the joint packet is determined by
the function (angular distribution function)
\begin{equation}
F_{I'}(\theta)=\frac{1}{4\pi}\left(1\pm
\frac{M_F}{I'+1}\cos\theta\right)\quad {\rm for}\ I'=I\pm 1.
\end{equation}
Modulation is due to the polarization of a parent ion. For a
non-polarized ion we have $F_{I'}=1/4\pi$.

The Gaussian packet in Eq.~(\ref{paxis}) for fixed $r_\nu$ has the form
\begin{equation}\label{pakn}
\exp\left[\frac{(1-\alpha)^2}{D^2_R}\!\left(r_n-\frac{\alpha
r_\nu}{1-\alpha}\right)^2\right],
\end{equation}
whereas for fixed $r_n$ it is equal
\begin{equation}\label{paknu}
\exp\left[\frac{\alpha^2}{D^2_R}\!\left(r_\nu-\frac{1-\alpha}
{\alpha}r_n\right)^2\right].
\end{equation}
Relative location of the peaks of these curves is determined by the
condition
\begin{equation}\label{cond}
(1-\alpha)r_n-\alpha r_\nu=0,
\end{equation}
which corresponds to the maximum of the wave packet (\ref{paxis}).
The $r_\nu$ dependent exponential and Gaussian curves overlap each
other, if $r_n$ does not exceed its maximum value $v_nt$.

\noindent
\section{Experimental implementations}

Recent experiments concerned with atom-photon entanglement are
dealing with generation and verification of an entangled pair. The
special structure of the recoil-neutrino wave packet offers new
kinds of experiments. There are three types of possible
experiments to detect a flavor neutrino and a recoil:
\vspace{-1.5mm}
\begin{itemize}
\item[(i)] Coincidence measurements, in which both the recoil
and neutrino are detected. The flavor-changing process
$e\rightarrow\beta$ is determined by the probability density
\begin{equation}\label{enp}
\sum_{M'}\!\Bigl|\bigr<0\bigr|\sum_j\!c_{j\mathbf{k}_{j0}}
U_{\beta j}\Psi_{I'\!M'}(\mathbf{r}_n,\mathbf{r}_\nu,t) \Bigl|^2.
\end{equation}\vspace{-3mm}
\item[(ii)] Noncoincidence measurements, when only neutrino is
detected regardless of the recoil position.The flavor-changing
process $e\rightarrow\beta$ is determined by the probability
density
\begin{equation}\label{no1}
\int d\mathbf{r}_n
\sum_{M'}\!\Bigl|\bigr<0\bigr|\sum_j\!c_{j\mathbf{k}_{j0}}
U_{\beta j}\Psi_{I'\!M'}(\mathbf{r}_n,\mathbf{r}_\nu,t) \Bigl|^2.
\end{equation}\vspace{-3mm}
\item[(iii)] Noncoincidence measurements, when only the recoil
nucleus is detected regardless of the neutrino position. The
probability density to detect recoil nucleus at the point
$\mathbf{r}_n$ is
\begin{equation}\label{no2}
\int d\mathbf{r}_\nu
\sum_{M'}\!\Bigl|\bigr<0\bigr|\sum_j\!c_{j\mathbf{k}_{j0}}
\Psi_{I'\!M'}(\mathbf{r}_n,\mathbf{r}_\nu,t) \Bigl|^2.
\end{equation}
\end{itemize}\vspace{-2mm}
The probability distributions (\ref{no1}) and (\ref{no2}) reveal
no entanglement because all information about the position of one
of the particles is lost completely.

{\it Coincidence measurement.} On the axis with highest
entanglement, the probability density to detect recoil nucleus at
the point $r_n$ together with the neutrino of flavor $\beta$ at
the point $r_\nu$ is
$$
\frac{dP_{e\beta}}{dr_ndr_\nu}=
F_{I'}(\theta)\Psi^2_{CM}\bigl((1-\alpha)r_n+\alpha
r_\nu,t\bigr)r^2_nr^2_\nu
$$
\vspace{-4mm}
\begin{equation}\label{prob}
\times\sum_{i,j}U_{\beta i}U^*_{ei} U_{ej}U^*_{\beta
j}R^*_i(r_\nu+r_n)R_j(r_\nu+r_n).\vspace{-2mm}
\end{equation}
In ultrarelativistic approximation the second line is the
probability of the $e\rightarrow\beta$ transition \vspace{-2mm}
\begin{equation}\label{osc}
\sum_i|U_{ei}|^2|U_{\beta i}|^2+2\sum_{i>j}|U_{\beta
i}U^*_{ei}U_{ej}U^*_{\beta j}|\cos\!\left(\frac{\Delta
m^2_{ij}}{2Q_{EC}}r\!+\!\gamma\!\right),
\end{equation}
where $\Delta m^2_{ij}=m^2_i-m^2_j$ and $\gamma={\rm arg}(U_{\beta
i}U^*_{ei}U_{ej}U^)_{\beta j})$. \vspace{1mm}

We choose the direction of the axis corresponding to the maximum
value of $F_{I'}(\theta)$, which will be denoted by $F_{I'}$. The
neutrino detector N, which fixes the flavor of neutrino states by
some charged current process, and the recoil one R to register the
nucleus are connected with a coincidence circuit and located on
this axis on both sides of a parent ion confining volume (that is
approximately the CM position) at distances from it, respectively,
$L_\nu$ and $L_n$. If $L_\nu$ and $L_n$ satisfy Eq.~(\ref{cond}),
the amplitude of oscillations will be equal
\begin{equation}\label{ampl}
\frac{2F_{I'}L^2_n}{\pi^{3/2}DD^3_R(t)}
\exp[-2\Gamma(t-L_\nu-L_n)].
\end{equation}
The amplitude conforms to highest spatial entanglement and can
serve as its verification. The detection time $t$ depends on the
lifetime of a parent ion. For short lifetimes, the length of the
joint wave packet may be equal to distance between the detectors,
that is $t=L_\nu$+$L_n$. In this case, the amplitude (\ref{ampl})
will be maximum. This condition is impossible for parent ions in a
Penning trap because their lifetime must exceed the time it takes
to prepare them in the trap. The latter is of the order of
seconds. In such a case, the measurement time $t>1/\Gamma$ should
be taken so as to minimize the exponent in Eq.~(\ref{ampl}).

It follows from Eq.~(\ref{osc}) that the expression for the
probability (\ref{prob}) contains the well-known oscillation phase
\begin{equation}
\phi_{ij}=\frac{\Delta m^2_{ij}}{2Q_{EC}}(L_\nu+L_n).
\end{equation}
For two neutrino mixing, i.e. in the case of $e\rightarrow\mu$
(i.e. $\beta=e,\mu$) oscillations, the mixing matrix is
\begin{equation}
U=\left(\begin{array}{cc}
\cos\eta &\sin\eta\\
-\sin\eta &\cos\eta
\end{array}\right),
\end{equation}
where $\eta$ is the mixing angle. Then, one finds from
Eq.~(\ref{osc}) the probabilities
\begin{equation}
{\cal W}_{ee}=1\!-\!\sin^2\!(2\eta)\sin^2\!\phi_{12},\ \ {\cal
W}_{e\mu}= \sin^2\!(2\eta)\sin^2\!\phi_{12},
\end{equation}
where $\phi_{12}=\pi(L_\nu+L_n)/L_0$ and $L_0\!=\!4\pi
Q_{EC}/\Delta m^2$ ($\Delta m^2=m^2_2-m^2_1>0$) is the oscillation
length. The final expression for the coincidence probability
density is
$$
\frac{dP_{e\beta}}{dr_nd\Omega_ndr_\nu d\Omega_\nu}\!=\!
\frac{2F_{I'}L^2_n}{\pi^{3/2}DD^3_R(t)}\!
$$
\vspace{-5mm}
\begin{equation}
 \times\exp[-2\Gamma(t-L_\nu\!-\!L_n)]{\cal W}_{e\beta}{\cal W}_n,
\end{equation}
where ${\cal W}_n=1$ is the probability of a recoil registration.
For the detectors connected to coincidence circuit with one
output (the "and circuit"), a signal disappearance at certain
positions of the detectors N and R is caused by oscillating
probability ${\cal W}_{e\mu}$ in the muon neutrino channel. The
recoil nucleus does not oscillate. To observe the spatial
correlation and neutrino oscillations, one needs the correlation
experiment, in which an experimental event involves the
registration of a neutrino simultaneously with a recoil nucleus in
two detectors. One can fix the distance from source to one of the
detectors and change the distance to another one. Here, too, both
neutrino oscillations and a recoil may be observed, but with a far
lower amplitude due to the Gaussian factor of Eq.~(\ref{paxis}).

Let us examine the conditions for the observation of oscillation
patterns. The first condition is trivial: the source-detector
distance $L_\nu$ should be of the order of, or greater than, the
oscillation lengths $L_{osc}$. The second condition is the
coherence of different mass eigenstates, necessary for the
neutrino and recoil oscillations to be observed. Coherence is
preserved over distances not exceeding the coherence length
$L_{coh}$. The latter is defined as the distance at which the
phase difference due to energy spreading obeys the equation
\vspace{-3mm}
\begin{equation}
\phi_{ij}(Q_{EC})-\phi_{ij}(Q_{EC}+\Gamma)=2\pi.
\end{equation}
We find $L_{coh}=L_{osc}Q_{EC}/\Gamma\sim 10^{21}L_{osc}$. Such a
large correlation length arises because the emitter is in a pure
quantum state that is described by a state vector (\ref{anz}).

{\it Neutrino detection.} This is a well-established experiment
for studying neutrino oscillation. Integration over
$d\mathbf{r}_n$ in Eq.~(\ref{no1}) can be done approximately since
the width of the Gaussian packet (\ref{pakn}) is much smaller than
that of the exponential, if $\Gamma t\ll10^{10}$. Indeed, we have
\begin{equation}
\frac{D_R}{(1\!-\!\alpha)D}\approx\frac{1}{D}
\sqrt{d^2\!+\!\frac{t^2}{M^2d^2}}\approx\frac{\Gamma t}{2Md}\sim
10^{-10}\Gamma t.
\end{equation}
The integral is calculated as follows
$$
\int\Psi^2_{CM}\bigl((1-\alpha)\mathbf{r}_n\!+\!\alpha\mathbf{r}_\nu\bigl)
R^*_i(|\mathbf{r}_\nu-\mathbf{r}_n|)R_j(|\mathbf{r}_\nu-\mathbf{r}_n|)d\mathbf{r}_n
 \vspace{-4mm}
$$
\begin{equation}
\approx R^*_i\!\left(\frac{r_\nu}{1-\alpha}\right)
R_j\!\left(\frac{r_\nu}{1\!-\!\alpha}\right)\!
\int\!\Psi^2_{CM}\bigl((1\!-\!\alpha)\mathbf{r}_n\bigr)d\mathbf{r}_n.
\end{equation}
Let the position of a neutrino detector be given by the distance
$L_\nu$ from the emitter, then the probability density to detect
the neutrino of flavor $\beta$ at this distance is
$$
\frac{dP_{e\beta}}{dr_\nu d\Omega_\nu}\!=\!
\frac{2F_{I'}}{D}\exp[-2\Gamma(t\!-\!L_\nu)]
\biggl\{\!\sum_i\!|U_{ei}|^2|U_{\beta i}|^2 \vspace{-4mm}
$$
\begin{equation}
+2\sum_{i>j}\!|U_{\beta i}U^*_{ei}U_{ej}U^*_{\beta
j}|\cos\biggl(\!2\pi
\frac{L_\nu}{L^\nu_{ij}}\!+\!\gamma\!\biggr)\!\biggr\}.
\end{equation}

{\it Recoil nucleus detection.} Integral (\ref{no2}) is calculated
in perfect analogy to (\ref{no1}) if $\Gamma t\ll10^5$ As a
result, we obtain the following expression for the probability
density to detect recoil at a distance $L_n=v_nt$ from the parent
ion
\begin{equation}
\frac{dP_n}{dr_n d\Omega_n}\!=\!\frac{F_{I'}\Gamma}{v_n}.
\end{equation}

The joint wave function of the recoil-neutrino pair allows to
calculate probability densities of the considered processes. For
full description of oscillation experiments, it is necessary to
take into account a detection process. However, this is different
problem.

\noindent
\section{Conclusion}

In summary, we have found an accurate analytical solution for the
joint quantum state of an electron neutrino and a recoil nucleus
following the electron capture decay of a hydrogenlike ion. The
evolution of the state provides an exactly calculable illustration
of the famous Einstein-Podolsky-Rosen thought experiment. The new
effect is the entanglement between the recoil and the coherent
superposition of the three massive neutrinos. Such a superposition
cannot be observed in the experiments involving discrete states of
an atom-photon pair because the polarization of an emitted photon
is inevitably entangled with the spin state of an atom.

We have shown that each of the three massive neutrinos is not
emitted in a momentum eigenstate, and so the recoil becomes
entangled with the superposition of three neutrinos. However, the
total energy is conserved in the course of EC decay, and the decay
energy converts to the kinetic energy of the reoil-neutrino pair.
Thus, we have rigorously proved that the neutrino mass eigenstates
composing the electron eigenstate produced in EC decay have the
same energy. It should be emphasized that although a particular
type of decay has been treated, the results of our calculations
are applicable to other two-body weak decays due to the common
features of the initial-value problem for noninteracting
particles.

The most peculiar aspect of our solution is the two-particle
system in a pure quantum state, where the particles, a recoil and
a massive neutrino, are mixed with each other. Such an
entanglement opens up an opportunity for correlative experiments
and the investigation of environment-induced decoherence. This is
the main distinction of our approach from the field-theoretical
one \cite{Grim1}, where a neutrino enters as unobserved
intermediate state. We have suggested the correlative experiment
which allows to observe neutrino oscillations and a recoil
simultaneously. However, a considerable progress in the detection
methods of neutrinos and recoil nuclei is necessary to carry out
such measurements with a single ion.

\noindent
\acknowledgments

The author is grateful to A. L. Barabanov and Y. Litvinov for
critical comments. This research was supported by Grant
NS-7235-2010.2 from the Russian Ministry of Education and Science.
\smallskip

\noindent
\appendix*
\section{}
We consider the integral
\begin{equation}\label{integ}
{\cal I}=\!\int\limits^\infty_0\frac{k_i(\epsilon)}{v_i(\epsilon)}
\frac{\exp{\!\left\{ik_i(\epsilon)r\!-\!
i\left[\epsilon\!+\!\frac{k^2_i(\epsilon)}{2M}\!
-\!\frac{\mathbf{p}\mathbf{k}_i(\epsilon)}{M}\right]\!t\right\}}}
{\epsilon\!-\!Q_{EC}\!+\!\frac{k^2_i(\epsilon)}{2M}\!
-\!\frac{\mathbf{p}\mathbf{k}_i(\epsilon)}{M}+i\Gamma}d\epsilon,
\end{equation}
where the function $k_i(\epsilon)$ is given by power series
(\ref{exp}). In the lowest orders of the small parameters $\alpha$
and for $\Gamma\ll Q_{EC}$ the pole of the integrand is
\begin{equation}\label{pole}
\epsilon_p\!=\!Q_{EC}\!-\!i\Gamma+\!\left(\frac{\mathbf{p}\mathbf{k}_{i0}}{M}\!-
\!\frac{k^2_{i0}}{2M}\right)\!\!\left(1\!+\!\frac{\mathbf{p}\mathbf{n}}{Mv_i}
\!-\!\frac{k_{i0}}{Mv_i}\right).
\end{equation}
We may extend the lower limit in (\ref{integ}) to $-\infty$, since
the contribution to the integral falls of sharply with increasing
$|\epsilon_i|$ owing to $\Gamma\ll Q_{EC}$. The integral involves
the exponential factor
$\exp{\!\left[-iq_i(\epsilon\!-\!Q_{EC})^2\right]}$ with
$$q_i\!=\!\frac{1}{2}\left[\frac{t}{Mv^2_i}\!+\!\frac{\delta^3_i}{m_iv^3_i}
\left(r\!+\!\frac{\mathbf{p}\mathbf{n}}{M}t\!-\!\frac{k_{i0}}{M}t\right)\right]
$$
 \vspace{-4mm}
\begin{equation}
\approx\frac{1}{2}\left(\frac{t}{Mv^2_i}\!+\!
\frac{\delta^3_ir}{m_iv^3_i}\right),
\end{equation}
which prevents the use of the residue method. To overcome the
problem, we have to use the Fourier transformation of this factor
\cite{Fed}
\begin{equation}
\exp{[-iq\varepsilon^2]}=\frac{1}{\sqrt{4i\pi
q}}\int\limits^\infty_{-\infty}\!\!\exp{\left(\frac{ix^2}{4q}+
i\varepsilon x\right)}dx.
\end{equation}

Then, we first evaluate the integral
\begin{equation}
\oint\limits_C\frac{k_i(\epsilon)\exp[i{\cal
P}(\epsilon)+i(\epsilon-Q_{EC})x]}{v_i(\epsilon)(\epsilon
-\epsilon_p)}d\epsilon,
\end{equation}
where the contour $C$ encloses the pole (\ref{pole}), and the
exponent ${\cal P}$ is
$$\hspace{-10mm} {\cal P}(\epsilon)=k_{i0}r-\left(Q_{EC}+\frac{k^2_{i0}}{2M}-
\frac{\mathbf{p}\mathbf{k}_{i0}}{M}\right)t
$$
 \vspace{-4mm}
\begin{equation}
+\left[\frac{r}{v_i}-\left(1-\frac{\mathbf{p}\mathbf{n}}{Mv_i}+
\frac{k_{i0}}{Mv_i}\right)t\right](\epsilon-Q_{EC}).
\end{equation}
The integral is easily calculated, and Eq.~(\ref{integ}) takes the
form \vspace{-1mm}
\begin{equation}\label{integ1}
{\cal I}=-\frac{2i\pi k_{i0}e^{i\cal P}}{\sqrt{4i\pi q_i}\,v_i}
\!\!\!\int\limits^\infty_{r/v_i-t}\!\!\!\!
\exp{\left[\frac{ix^2}{4q_i}\!+\!
ix(\epsilon_p\!-\!Q_{EC})\right]}dx,
\end{equation}
where for $\Gamma\ll Q_{EC}$ the expression for the exponent taken
in the pole has the form \vspace{-2mm}
$$
{\cal P}\!=\!{\cal P}(\epsilon_p)\!= \!
k_{i0}r\left(1\!-\!\frac{\alpha}{2}\!+\!
\frac{\alpha^2}{2}\right)\!-\!
Q_{EC}\!\left(1\!+\!\frac{\alpha^3}{2}\!-\!\frac{\alpha
m^2_i}{2M^2}\right)t
$$
 \vspace{-4mm}
$$
+\,\alpha\mathbf{p}\mathbf{r}\left(1\!-\!\frac{3\alpha}{2}\!+\!
2\alpha^2\frac{v_it}{r}\right)+
i\Gamma\left[(1\!-\!\alpha^2)t\!-\!\frac{r}{v_i}(1\!-\!\alpha)\right]
$$
 \vspace{-4mm}
\begin{equation}\label{expn}
+\alpha\frac{(\mathbf{p}\mathbf{n})^2}{M}\left(\frac{r}{v_i}\!-
\!\frac{5\alpha}{2}t\right)+\alpha\frac{(\mathbf{p}\mathbf{n})^3}{M^2v_i}t.
\end{equation}
We have approximately
\begin{equation}
{\cal P}\approx k_{i0}r+\alpha\mathbf{p}\mathbf{r}-
Q_{EC}t+i\Gamma(t-r/v_i).
\end{equation}
The last two terms in the expression (\ref{expn}) can be dropped
because they are small compared to the term proportional to
$\mathbf{p}^2$ in the exponent of Eq.~(\ref{psi}). It is seen from
Eq.~(\ref{expn}) that the energy of a pair of $i$th neutrino and a
recoil is the same for all pairs with great accuracy and is equal
to the decay energy $Q_{EC}$. Small deviations from $Q_{EC}$ are
attributable to the approximate calculation of the pole
(\ref{pole}): the more precisely it is defined, the less are these
deviations.

Finally, the integral (\ref{integ1}) can be transformed into the
error function
\begin{equation}
{\cal I}=-\frac{i\pi k_{i0}}{v_i}\left(1-{\rm
Erf}\frac{r/v_i-t+iq_i\Gamma}{\sqrt{4iq_i}}\right)e^{i\cal P}.
\end{equation}


\begin{thebibliography}{99}
\bibitem{vol} The volume of a normalized box will be taken 1
throughout this paper.
\end{thebibliography}
\end{document}